\documentclass{DISproc}

\DeclareMathOperator{\tr}{Tr}

\renewcommand{\d}{\mathrm{d}}
\newcommand{\sT}{{\scriptscriptstyle T}}
\newcommand{\bm}[1]{\mbox{\boldmath $#1$}}
\newcommand{\pT}{\mathbf{p}_\sT}
\newcommand{\kT}{\mathbf{k}_\sT}

\begin{document}
\title{Linearly polarized Gluons and the Higgs\\ Transverse Momentum Distribution}

\author{{\slshape Wilco J. den Dunnen$^1$, Dani\"el Boer$^2$, Cristian Pisano$^3$, 
Marc Schlegel$^4$, Werner Vogelsang$^4$}\\[1ex]
$^1$ 
  VU University Amsterdam, NL-1081 HV Amsterdam, The Netherlands\\
$^2$ 
 KVI, University of Groningen, Zernikelaan 25, NL-9747 AA Groningen, The Netherlands\\
$^3$ 
 INFN, Sezione di Cagliari, C.P. 170, I-09042 Monserrato (CA), Italy\\
$^4$ 
     Universit\"{a}t T\"{u}bingen,
     Auf der Morgenstelle 14,
     D-72076 T\"{u}bingen, Germany}

\contribID{51}

\doi  

\maketitle

\begin{abstract}
We investigate the possible role of linearly polarized gluons in Higgs 
production from unpolarized $pp$ collisions. 
The transverse momentum distribution of the produced 
Higgs boson is found to exhibit a modulation with respect to 
the naive, unpolarized expectation, with the sign depending on the parity
of the Higgs boson. 
The transverse momentum distribution of a scalar Higgs will, therefore, have a 
shape clearly different from a pseudo-scalar Higgs. 
We suggest that this effect can be used to determine the parity of the 
Higgs at the LHC, without the need to use challenging angular distributions.
\end{abstract}

\subsection*{Introduction}

After a discovery of a new scalar particle at the LHC, the next task at hand is to
determine its coupling to other particles. 
Not only the size, but also the type of coupling to fermions has to be determined,
being either the $P$ even $\overline\Psi \Psi$ or the $P$ odd $\overline\Psi\gamma_5\Psi$.
Relatively few suggestions to this end have been put forward 
for the LHC, e.g., using Higgs + 2 jet production 
\cite{ref:Higgs2jet}
or $\tau$ pair decays 
\cite{ref:Higgstaudecay}.
We claim that the difference between a scalar and pseudoscalar coupling might
also be visible in the transverse momentum distribution of the scalar particle.

The Higgs transverse momentum distribution has been calculated in the
framework of collinear factorization at Next-to-Next-to-Leading Logarithmic
(NNLL) accuracy for small $q_\sT$, matched to Next-to-Leading Order (NLO) accuracy 
for large $q_\sT$ 
\cite{ref:Bozzis,Balazs:2007hr}.
It was noted \cite{Balazs:2007hr,Nadolsky:2007ba} that in NLO $\gamma\gamma$ continuum
production there are ``gluon spin-flip contributions'' in the $gg$ induced channel,
which should be described by a ``spin-flip distribution'' $\mathcal{P}^\prime$, that
can, in principle, be as large as the unpolarized distribution. 
It was also noted \cite{Catani:2010pd} that in the NNLO radiative corrections to the Higgs
boson $q_\sT$ cross section ``gluon spin correlations'' become important, which cause
the standard Drell-Yan transverse momentum resummation to fail for 
the gluon-gluon fusion process.
We claim that the fact that these gluon polarization effects are only observed at NNLO in 
Higgs production, is due to the use of the collinear factorization framework in which 
the polarized gluons have to be generated from the unpolarized distribution by gluon radiation.
Within the framework of Transverse Momentum Dependent (TMD) factorization, the effect of 
polarized gluons is already present at tree-level and described by a non-perturbative
input function $h_1^{\perp g}$.
Although dependent on its size, the effects of polarized gluons are,
in principle, large and modify the $q_\sT$ distribution of a scalar and pseudoscalar Higgs
in a distinct way.

\subsection*{TMD factorization}

The contribution of gluon fusion to Higgs production in the TMD framework reads \cite{Boer:2011kf} 
in leading order in $q_\sT/m_H$,
\begin{equation}\label{eq:TMDfactorization}
\frac{\d\sigma}{\d^{3}\vec{q}} 
\propto 
\int\! \d^{2}\pT\, \d^{2}\kT\,
\delta^{2}(\pT+\kT-\bm q_{\sT})\,
\Phi_g^{\mu \nu}(x_{a},\,\pT)
\Phi_g^{\rho \sigma}(x_{b},\,\kT)
\left(\hat{\mathcal{M}}^{\mu \rho}\right)\left(\hat{\mathcal{M}}^{\nu \sigma}\right)^{\ast}
\Big|_{p_{a}=x_{a}P_{a}}^{p_{b}=x_{b}P_{b}},
\end{equation} 
in which the momentum fractions are given by $x_{a(b)}=q^2/(2P_{a(b)}\cdot q)$
and $\Phi_g$ is the gluon correlator \cite{Mulders:2000sh,Boer:2010zf},
\begin{align}\label{eq:gluoncorrelator}
\Phi_g^{\mu\nu}(x,\bm p_\sT )
&=  	\frac{n_\rho\,n_\sigma}{(p{\cdot}n)^2}
	{\int}\frac{d(\xi{\cdot}P)\,d^2\xi_\sT}{(2\pi)^3}\
	e^{ip\cdot\xi}\,
	\langle P|\,\tr\big[\,F^{\mu\rho}(0)\,
	F^{\nu\sigma}(\xi)\,\big]
	\,|P \rangle\,\big\rfloor_{\text{LF}}\nonumber\\
&=
	-\frac{1}{2x}\,\bigg \{g_\sT^{\mu\nu}\, f_1^g(x,\bm{p}_\sT^2)
	-\bigg(\frac{p_\sT^\mu p_\sT^\nu}{M^2}\,
	{+}\,g_\sT^{\mu\nu}\frac{\bm p_\sT^2}{2M^2}\bigg)
	\; h_1^{\perp\,g}(x,\bm{p}_\sT^2) \bigg \} + \text{higher twist},
\end{align}
with $p_{\sT}^2 = -\bm p_{\sT}^2$, $g^{\mu\nu}_{\sT} = g^{\mu\nu}
- P^{\mu}n^{\nu}/P{\cdot}n-n^{\mu}P^{\nu}/P{\cdot}n$, and $M$ the proton mass.
The function $f_1^g(x,\bm{p}_\sT^2)$ represents the unpolarized gluon distribution 
and $h_1^{\perp\,g}(x,\bm{p}_\sT^2)$ represents the distribution of linearly polarized gluons.

\subsection*{Size of the linearly polarized gluon distribution}

As a first step, to study the effects of linearly polarized gluons,
we follow a standard approach for TMDs in the literature 
and assume a simple Gaussian dependence of the gluon TMDs on transverse momentum:
\begin{equation}\label{eq:f1par}
f_1^g(x,\bm p_\sT^2) = \frac{G(x)}{\pi \langle  p_\sT^2 \rangle}\,
\exp\left(-\frac{\bm p_\sT^2}{\langle  p_\sT^2 \rangle}\right),
\end{equation}
where $G(x)$ is the collinear gluon distribution,
$G(x)= \int \d^2\pT\, f_1^g(x,\pT)$.
The width, $\langle  p_\sT^2 \rangle$, depends on the energy scale, $Q$, and should be 
experimentally determined.
We will estimate $\langle p_\sT^2 \rangle = 7\,\mathrm{GeV}^2$,
at $Q=m_H=120\, \text{GeV}$, in rough agreement with the Gaussian fit to 
$f_1^u(x, \bm p_\sT^2)$ evolved to $Q=M_Z$ of Ref.\ \cite{Aybat:2011zv}.

No experimental data on $h_1^{\perp g}$ is available,
but a positivity bound has been derived in Ref.~\cite{Mulders:2000sh}:
$\frac{\bm p_\sT^2}{2M^2}\,|h_1^{\perp g}(x,\bm p_\sT^2)|\le f_1^g(x,\bm p_\sT^2)$.
We will use a Gaussian Ansatz for $h_1^{\perp g}$, with a width of 
$r \langle p_\sT^2 \rangle$,
\begin{equation}\label{eq:h1ppar}
h_1^{\perp g}(x,\bm p_\sT^2)=\frac{M^2G(x)}{\pi \langle p_\sT^2 \rangle^2}\frac{2e(1-r)}{r}\,
\exp\left(-\frac{\bm p_\sT^2}{r \langle p_\sT^2 \rangle}\right),
\end{equation}
with a normalization such that it satisfies the bound for all $\bm p_\sT$.

\subsection*{Higgs transverse momentum distribution}

\begin{figure}[htb]
\centering
\includegraphics[width=0.49\textwidth]{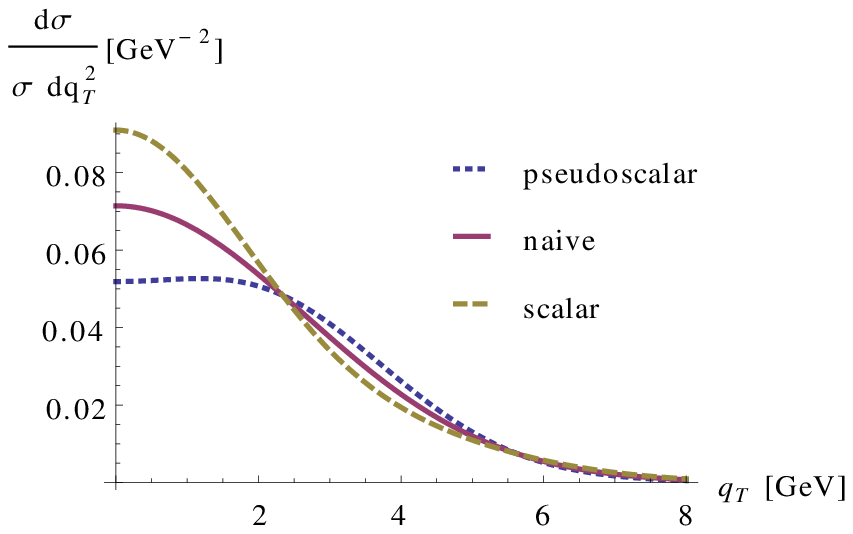}
\includegraphics[width=0.49\textwidth]{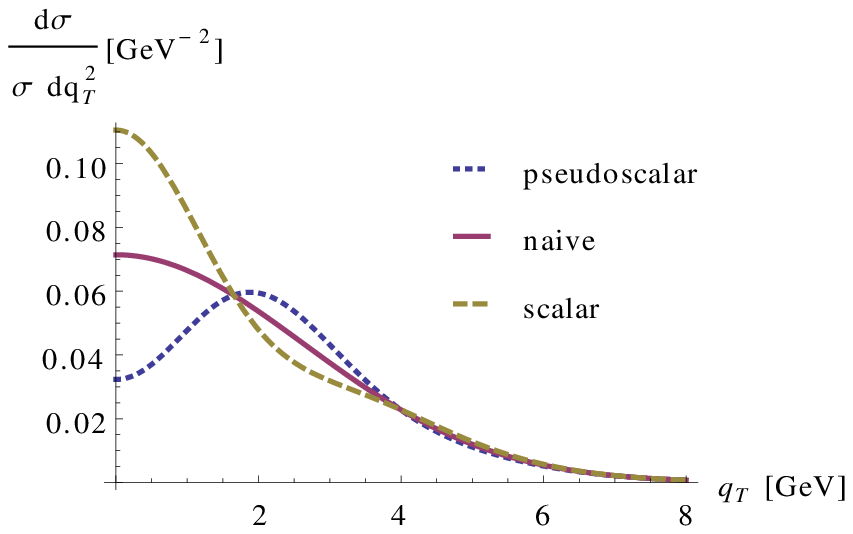}
\caption{Transverse momentum distribution of the Higgs, using the parameterization of 
$h_1^{\perp g}$ in Eq.\ \eqref{eq:h1ppar} with $r=2/3$ (left) and $r=1/3$ (right). 
The naive curve is the prediction for both scalar and pseudoscalar in case $h_1^{\perp g}=0$.}
\label{fig:qTdistr}
\end{figure}

Using the TMD factorization expression in Eq.\ \eqref{eq:TMDfactorization},
the parameterization of the gluon correlator in Eq.\ \eqref{eq:gluoncorrelator}
and the Ansatz for the TMD distribution in Eqs.\ \eqref{eq:f1par} and \eqref{eq:h1ppar},
the transverse momentum distribution for a scalar and pseudoscalar Higgs can be written as
\begin{equation}\label{eq:qTdistr}
\frac{1}{\sigma}\frac{\d\sigma}{\d q_\sT^2} = \left[1 \pm R (q_\sT)\right] \frac{1}{2 \langle p_\sT^2 \rangle} 
	e^{-q_\sT^2/ 2\langle p_\sT^2 \rangle},
\end{equation}
where $\pm$ stands for a scalar/pseudoscalar and 
$R (q_\sT) \equiv \mathcal{C}[w_H h_1^{\perp g} h_1^{\perp g}]/\mathcal{C}[f_1^g f_1^g]$,
with
\begin{equation}
\mathcal{C}[w\, f\, f] \equiv \int \d^{2}\pT\int \d^{2}\kT\,
\delta^{2}(\pT+\kT-\bm q_{\sT})
w(\pT,\kT)\, f(x_{a},\pT^{2})\, f(x_{b},\kT^{2})
\end{equation}
and $w_{H}=[(\pT \cdot \kT)^{2}-\frac{1}{2}\pT^2 \kT^2]/2M^{4}$.
With our Ansatz for the distribution functions,
\begin{equation}
R (q_\sT) = \frac{r}{2} (1-r)^2 \left(1 - \frac{q_\sT^2}{r\, \langle p_\sT^2 \rangle} 
	+ \frac{q_\sT^4}{8\, r^2\, \langle p_\sT^2 \rangle^2 }\right) 
	\exp\left[2-\frac{1-r}{r}\frac{q_\sT^2}{2\, \langle p_\sT^2 \rangle}\right].
\end{equation}
The transverse momentum distribution in Eq.\ \eqref{eq:qTdistr} for a scalar and
pseudoscalar Higgs is plotted in Figure \ref{fig:qTdistr} for $r=2/3$ and $r=1/3$.
As long as $h_1^{\perp g}$ is not measured, the absolute size of the effect is unknown,
but it will always be such that a scalar has enhancement at low $q_\sT$, suppression
at moderate $q_\sT$, followed again by enhancement at high $q_\sT$, whereas for the pseudoscalar
this is reversed.
Higher order perturbative corrections will modify the
exact form and width of our tree-level $q_\sT$ distribution, as well
as the size of the modulation, but this qualitative behavior is expected not to change.

\vspace*{-0.3cm}
\subsection*{Two photon decay channel}

\begin{figure}[htb]
\centering
\includegraphics[width=0.49\textwidth]{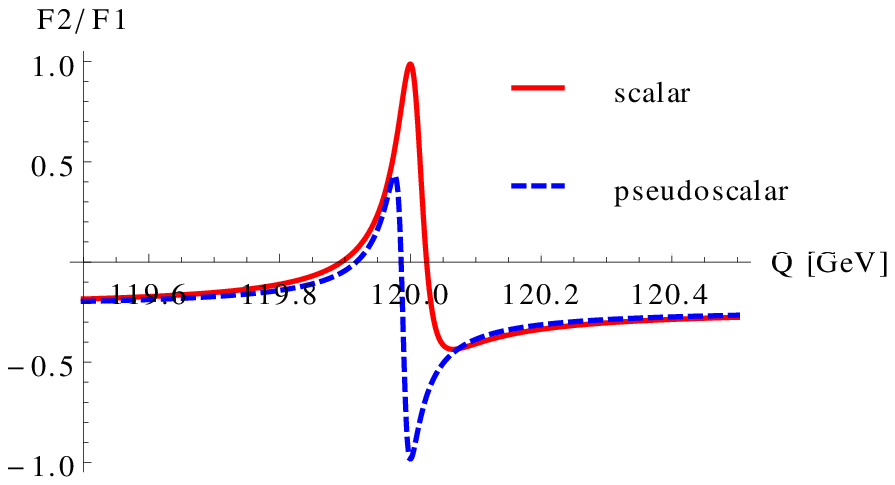}
\includegraphics[width=0.49\textwidth]{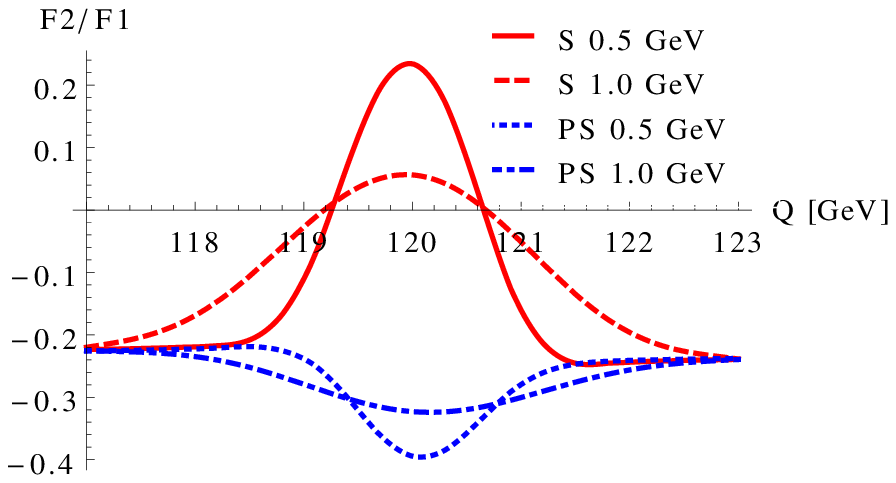}
\caption{The ratio $F_2/F_1$ in Eq.\ \eqref{eq:qTgammagamma}
plotted as function of $Q$ for $\theta=\pi/2$ assuming a $120$\,GeV Higgs (left) and the same curves including a 
detector resolution of 0.5 and 1\,GeV (right).}
\label{fig:F2F1}
\end{figure}

When the Higgs decays to, e.g., two photons, there will
be irreducible background due to $gg\to\text{quark box}\to \gamma\gamma$, which
was recently investigated in the framework of TMD factorization \cite{Qiu:2011ai}.
Including this background, we come to the conclusion \cite{Boer:2011kf}, 
that the transverse momentum distribution of the 
photon pair has the same form as Eq.\ \eqref{eq:qTdistr}, 
but with a $Q$ and Collins-Soper angle, $\theta$,
dependent size, i.e.,
\begin{equation}\label{eq:qTgammagamma}
\frac{1}{\sigma}\frac{\d\sigma}{\d q_\sT^2} = \left[1 + \frac{F_2(Q,\theta)}{F_1(Q,\theta)} R (q_\sT)\right] \frac{1}{2 \langle p_\sT^2 \rangle} 
	e^{-q_\sT^2/ 2\langle p_\sT^2 \rangle}.
\end{equation}
The ratio $F_2/F_1$ is plotted in the left graph of Fig.\ \ref{fig:F2F1}.
At the Higgs mass we reproduce Eq.\ \eqref{eq:qTdistr}, i.e.\ $F_2/F1\to \pm 1$, 
for a scalar/pseudoscalar, but away from the pole, the background quickly dominates.
To mimic a finite detector resolution in the determination of $Q$, 
we also plot the ratio $F_2/F_1$ in which both numerator and denominator are 
separately weighted with a Gaussian distribution.
From the graph we see that the continuum background reduces the effect to 
approximately 30\% or 20\% of the maximal size with a 0.5 or 1\,GeV 
resolution, respectively.

\subsection*{Conclusions}

The effect of gluon polarization in $pp$ collisions is such that
scalar and pseudoscalar particles, produced through gluon fusion,
will have different transverse momentum distributions.
Although the absolute size of the effect cannot be estimated
without experimental input on $h_1^{\perp g}$,
the qualitative features are such that this effect can,
in principle, be used to distinguish scalar from 
pseudoscalar particles.
In the two photon decay channel of the Higgs, 
the continuum background partially washes out the difference
between scalar and pseudoscalar.
Other decay channels are currently being investigated.


This work is part of the research program of the ``Stichting voor Fundamenteel Onderzoek der Materie (FOM)'' 
which is financially supported by the ``Nederlandse Organisatie voor Wetenschappelijk Onderzoek (NWO)''.


{\raggedright
\begin{footnotesize}



\end{footnotesize}
}


\end{document}